\def\mb#1{\mbox{\boldmath{$#1$}}} % italic math bold
\documentclass[showkeys,showpacs]{revtex4}
\usepackage{graphicx}
\begin{document}
\hspace*{4.5 in}CUQM-135
%\hspace*{3.5 in}susy.tex $28^{\rm st}$ May 2010
\vskip 0.4 in

%\begin{frontmatter}

%%%%%%%%%%%%%%%%%%%%% Publisher's Area please ignore %%%%%%%%%%%%%%%
%\catchline{}{}{}{}{}
%%%%%%%%%%%%%%%%%%%%%%%%%%%%%%%%%%%%%%%%%%%%%%%%%%%%%%%%%%%%%%%%%%%%

\title{Supersymmetric analysis for the Dirac equation with spin-symmetric and pseudo-spin-symmetric interactions}

\author{Richard~L.~Hall}

\author{\"Ozlem Ye\c{s}ilta\c{s}\footnote{Permanent address: Department of Physics, Faculty of Arts and Sciences,
Gazi University, 06500 Ankara, Turkey.}}
\address{Department of Mathematics and Statistics, Concordia
University, 1455 de Maisonneuve Boulevard West, Montr\'eal,
Qu\'ebec, Canada H3G 1M8}
\email{rhall@mathstat.concordia.ca}
\email{yesiltas@gazi.edu.tr}

\begin{abstract}
A supersymmetric analysis is  presented for the $d$-dimensional Dirac equation with central potentials under spin-symmetric ($S(r) = V(r)$) and pseudo-spin-symmetric ($S(r) = - V(r)$) regimes.  We construct the explicit shift operators that are required to factorize the Dirac Hamiltonian with the Kratzer potential. Exact solutions are provided for both the Coulomb and Kratzer potentials.
\end{abstract}

\maketitle
%\newpage
%%%%%%%%%%%%%%%%%%%%%%%%%%%%%%%%%%%%%%%%%%
\section{Introduction}
%%%%%%%%%%%%%%%%%%%%%%%%%%%%%%%%%%%%%%%%%%
Exact solutions of the Dirac equation that are allowed by certain soluble potentials have always been of interest in relativistic quantum theory \cite{che,jia,dong,hall,chen,berkdemir,dong1,dong2,lisboa}. More particularly, certain aspects of deformed nuclei have been studied for over thirty years  \cite{hecht,arima} by means of spin-symmetric and pseudo-spin-symmetric concepts. Ginocchio  \cite{ginoc,ginoc1,ginoc2} showed that spin symmetry occurs when the difference between the vector potential $V(r)$ and scalar potential $S(r)$ is a constant, $V(r)-S(r)={\rm const.}$; and pseudo-spin symmetry occurs when the sum of the vector potential $V(r)$ and scalar potential $S(r)$ is a constant, $V(r)+S(r)={\rm const.}$ In the spin-symmetric limit the Dirac equation corresponds to a Schr\"odinger equation which possesses SU(3) symmetry \cite{ginoc}. The pseudo-spin symmetry constraint implies a degeneracy of the single-nucleon doublets: this can be shown explicitly in terms of the non-relativistic quantum numbers $(n, \ell, j = \ell + 1/2)$ and $(n -1, \ell + 2, j = \ell + 3/2)$, where $n, \ell$ and $j$ are the single-nucleon radial, orbital, and total angular-momentum quantum numbers, respectively. 

In this paper, we formulate the spin-symmetric and pseudo-spin-symmetric problems generated by the Dirac equation in such a way that they are amenable to an analysis in terms of factorization and shape-invariance methods \cite{cooper}.  In particular, we provide a supersymmetric analysis of such Dirac Hamiltonians with Coulomb and Kratzer potentials. This leads to expressions for  all the bound states and corresponding energy eigenvalues for these spin-symmetric and pseudo-spin-symmetric problems. In particular we show that the spin and pseudo-spin symmetry limits lead respectively to distinctly different features for the ground state, which, to  our knowledge, have not been noted before. Some earlier exact solutions of the Dirac equation within the context of supersymmetric quantum mechanics may be found in Refs.~\cite{cs,roy,al}.

We consider a single particle that is bound by attractive central vector and scalar potentials,
respectively $V$ and $S$, in $d\ge 1$ spatial dimensions and obeys the Dirac equation.  For central potentials in $d$ dimensions the Dirac equation can be written \cite{jiang} in natural units $\hbar=c=1$ as
\begin{equation}\label{eq1}
i{{\partial \Psi}\over{\partial t}} =H\Psi,\quad {\rm where}\quad  H=\sum_{s=1}^{d}{\alpha_{s}p_{s}} + (m+S)\beta+V,
\end{equation}
$m$ is the mass of the particle, $V(r)$ and $S(r),$ $ r=|\mb{r}|,$ are the spherically symmetric vector and scalar potentials, and $\{\alpha_{s}\}$ and $\beta$  are Dirac matrices, which satisfy anti-commutation relations; the identity matrix is implied after the vector potential $V$. For stationary states, algebraic calculations in a suitable basis lead to a pair of first-order linear differential equations in two radial functions $\{\psi_1(r), \psi_2(r)\}$.  For $d > 1,$ these functions vanish at $r = 0$, and, for bound states, they may be normalized by the relation
\begin{equation}\label{eq2}
(\psi_1,\psi_1) + (\psi_2,\psi_2) = \int\limits_0^{\infty}(\psi_1^2(r) + \psi_2^2(r))dr = 1.
\end{equation}
We use inner products {\it without} the radial measure factor $r^{(d-1)}$ because the factor $r^{\frac{(d-1)}{2}}$ is already built in to each radial function. Thus the radial functions vanish at $r = 0$ and satisfy the coupled equations
\begin{eqnarray}
E\psi_1 &=& (V+m+S)\psi_1 + (-\partial + k_{d}/r)\psi_2\label{eq3}\\
E\psi_2 &=& (\partial + k_{d}/r)\psi_1 + (V-m-S)\psi_2\label{eq4},
\end{eqnarray}
where $k_1 = 0,$ $k_{d}=\tau(j+{{d-2}\over{2}}),~d >1$,  and $\tau = \pm 1$ and $\partial$ represents the operator $\partial/\partial r$. We note that the variable $\tau$ is sometimes written $\omega$, as, for example in the book by Messiah \cite{messiah}. The quantum number $k_d$ is related to $\ell$ and $j$ for the spin-symmetric and pseudo-spin-symmetric cases as follows:
\begin{eqnarray}
k_d=\left\{
  \begin{array}{ll}
    -(\ell+\frac{d-1}{2}), & \hbox{$j=\ell+\frac{1}{2}$;} \\
    \ell+\frac{d-3}{2}, & \hbox{$j=\ell-\frac{1}{2}$.}
  \end{array}
\right.
\end{eqnarray}
and
\begin{eqnarray}
k_d=\left\{
  \begin{array}{ll}
    (\tilde{\ell}+\frac{d-1}{2}), & \hbox{$j=\tilde{\ell}+\frac{1}{2}$;} \\
    -(\tilde{\ell}+\frac{d-3}{2}), & \hbox{$j=\tilde{\ell}-\frac{1}{2}$.}
  \end{array}
\right.
\end{eqnarray}
Here $\tilde{\ell}=\ell+1$ is called the pseudo orbital angular momentum \cite{ginoc2,hecht}. The radial functions are often written $\psi_1 = G$ and $\psi_2 = F,$ as in the book by Greiner \cite{greiner}.  We shall assume that the potentials $V$ and $S$ are such that there are some discrete eigenvalues $E_{k_d n}$ and that Eqs.(\ref{eq3}) are the eigenequations for the corresponding radial eigenstates. Here $n=0,1,2,\dots$ enumerates the radial wave functions for a given $k_d$.  In this paper we shall present the problem explicitly for the cases $d > 1.$ Similar arguments go through for the case $d=1$: in this case $k_1 = 0,$ the states can be classified as even or odd, and the normalization (\ref{eq2}) becomes instead $\int_{-\infty}^{\infty}\left(\psi_1^2(x) + \psi_2^2(x)\right)dx = 1.$
%%%%%%%%%%%%%%%%%%%%%%%%%%%%%%%%%%%%%%%%%%%%%%%%%%%%%%%%%%%%%
\section{Shape invariance and SUSY partner Hamiltonians}
%%%%%%%%%%%%%%%%%%%%%%%%%%%%%%%%%%%%%%%%%%%%%%%%%%%%%%%%%%%%%
A supersymmetric (SUSY) Hamiltonian satisfies the following graded Lie algebra \cite{witten}
\begin{eqnarray}
\mathcal{H}=\{Q, Q^{\dag}\},~~~~Q=\left(
  \begin{array}{cc}
    0 & 0 \\    A^{-}_{0} & 0 \\  \end{array}\right), ~~~~ Q^{\dag}=\left(\begin{array}{cc}  0 & A^{+}_{0} \\
                           0 & 0 \\    \end{array}  \right),\nonumber
\end{eqnarray}
where the supercharges $Q$ and $Q^{\dag}$ commute with $\mathcal{H}$ and are nilpotent operators: $(Q^{\dag})^{2}=0=Q^{2}$. Here, the operators $A^{\pm}_0$ are given by
\begin{equation}\nonumber
    A^{\pm}_0=\pm \frac{d}{dr}+W(r),
\end{equation}
where $W(r)$ is the superpotential. The associated SUSY partner Hamiltonians $H_{1}$ and $H_{2}$ have the standard
forms
\begin{equation}\nonumber
    H_{1}=-\frac{d^{2}}{dr^{2}}+V_{1}(r),~~~~H_{2}=-\frac{d^{2}}{dr^{2}}+V_{2}(r).
\end{equation}
In the case of unbroken SUSY, there is a remarkable aspect of SUSY-QM, namely, except for the zero-energy eigenstate, the SUSY partner Hamiltonians  $H_1$ and $H_2$ are found to be exactly isospectral. Here $V_{1}(r)$ and $V_{2}(r)$ are corresponding partner potentials satisfying
\begin{equation}\label{v12}
    V_{1}(r;a_{1})=V_{2}(r;a_{2})+R(a_{1}), ~~~~a_{2}=f(a_{1}).
\end{equation}
where $a_1$ and $a_2$ are constants and the remainder $R(a_1)$ is independent of $r$. Often the relation between the potentials is given by $V_{2}(r;a_{1})=V_{1}(r;a_{2})+R(a_{1})$ in the literature, but we prefer to use the notation of Sukumar \cite{sukumar}, which allows $\psi_1$ and $\psi_2$, respectively, to be retained as the upper and lower radial functions in the Dirac spinor. By using the shape invariance condition (\ref{v12}), the entire spectrum of $H_2$ can be found.
Thus one can construct the sequence of Hamiltonians by using iteration \cite{cooper}, to find
\begin{equation}\label{it}
    H_s=-\frac{d^{2}}{dr^{2}}+V_2(r;a_s)+\sum^{s-1}_{k=1} R(a_k),~~~a_s=f^{s-1}(a_1), ~~~s=1,2,...,
\end{equation}
where $f^s(a)$ means $s$ repeated applications $f(f(\dots(f(a))\dots)$ of the function $f$.
In this fashion, the entire spectrum of eigenenergies for the initial Hamiltonian $H_2$  can be obtained algebraically by
\begin{equation}\label{e}
    E^{(2)}_0=0, ~~~~ E^{(2)}_n=\sum^{n}_{k=1} R(a_{k}).
\end{equation}
Here the superscript $(2)$ denotes the eigenvalue for the Hamiltonian $H_2$. The corresponding eigenfunctions for the $H_1$ and $H_2$ are written simply $\psi_{1}$ and $\psi_{2}$, and also, more fully, for the corresponding $n^{\rm th}$ excited states, $\psi^{(1)}_n$ and $\psi^{(2)}_n$. The unnormalized energy eigenfunction $\psi^{(2)}_n$ for the Hamiltonian $H_2$ reads
\begin{equation}\label{unn}
    \psi^{(2)}_n \propto (A^{+}_0 A^{+}_1...A^{+}_{n-1})\psi^{(2)}_0(r;a_n+1)
\end{equation}
but one can also use \cite{cooper}
\begin{equation}\label{suk}
    \psi^{(2)}_n(r;a_1)=A^{+}_0(r;a_1) \psi^{(2)}_{n-1}(r;a_2).
\end{equation}
The complete spectrum is given by
\begin{equation}\label{hier}
    E^{(2)}_{n+1}=E^{(1)}_{n}.
\end{equation}
The reader may wish to look in Ref.~\cite{cooper} for more details.
%%%%%%%%%%%%%%%%%%%%%%%%%%%%%%%%%%%%%%%%%%%%%%%%%%%%%%%%%%%%%%%%%%%%%%%%%%%
\section{Factorization and shape invariance for the Coulomb Problem}
%%%%%%%%%%%%%%%%%%%%%%%%%%%%%%%%%%%%%%%%%%%%%%%%%%%%%%%%%%%%%%%%%%%%%%%%%%%
%%%%%%%%%%%%%%%%%%%%%%%%%%%%%%%%%%%%%%%%%%%%%%%%%
\subsection{Spin-symmetric problems }
%%%%%%%%%%%%%%%%%%%%%%%%%%%%%%%%%%%%%%%%%%%%%%%%%
In this case (\ref{eq1}) can be written in the form:
\begin{eqnarray}\label{m1}
H \left(   \begin{array}{c}     \psi_1 \\     \psi_2 \\   \end{array} \right)  =\left(
    \begin{array}{cc}      0 & m+E \\       m-E & 0 \\
    \end{array}  \right)  \left(    \begin{array}{c}       \psi_1 \\       \psi_2 \\    \end{array}  \right)
\end{eqnarray}
in which
\begin{equation}\label{mm}
H = \frac{d}{dr} \hat{1} + U,
\end{equation}
$\hat{1}$ is the identity matrix, and the matrix $U$ is given by
\begin{eqnarray}
\nonumber U =\left(   \begin{array}{cc}
     \frac{k_d}{r} & 0 \\      -2V(r) & -\frac{k_d}{r} \\   \end{array}  \right).
\end{eqnarray}
We shall now use the Coulomb potential $V(r)=-\frac{v}{r}$ in this matrix equation. We can diagonalize the matrix $U$ by means of the similarity transformation $D^{-1} U D$, where
\begin{eqnarray}
 D=\left(                \begin{array}{cc}
                  \frac{k_d}{v} & 0 \\                  1 & 1 \\                \end{array}
              \right). \nonumber
\end{eqnarray}
We define $\tilde{\psi_1}$ and $\tilde{\psi_2}$ as the transformed radial components obtained by
\begin{eqnarray}
\left(          \begin{array}{c}            \tilde{\psi}_1 \\
            \tilde{\psi}_2 \\           \end{array}         \right)=D^{-1}\left(
  \begin{array}{c}
    \psi_1 \\    \psi_2 \\  \end{array} \right). \nonumber
\end{eqnarray}
If we multiply (\ref{m1}) by $D^{-1}$ from the left, we have
 \begin{equation}\label{o1}
    A^{+}_{0} \tilde{\psi}_1=(m+E)\frac{v}{k_d}\tilde{\psi}_2
 \end{equation}
and
 \begin{equation}\label{o2}
    A^{-}_{0} \tilde{\psi}_2=\left((m+E)\frac{v}{k_d}-\frac{k_d}{v}(m-E)\right)\tilde{\psi}_1,
 \end{equation}
where the operators $A^{\pm}_{0}$ have the form:
\begin{equation}\label{op0}
    A^{\pm}_{0}=\pm \frac{d}{dr}+\frac{k_d}{r}-(m+E)\frac{v}{k_d}.
\end{equation}
Thus we may now write the eigenvalue equations for $\tilde{\psi_1}$ and $\tilde{\psi_2}$, namely
\begin{eqnarray}\label{E1}
% \nonumber to remove numbering (before each equation)
  A^{-}_{0} A^{+}_{0} \tilde{\psi_1} &=& \varepsilon \tilde{\psi_1} \\
  A^{+}_{0} A^{-}_{0} \tilde{\psi_2} &=& \varepsilon \tilde{\psi_2}, \nonumber
\end{eqnarray}
  where the eigenvalue $\varepsilon$ is given by
\begin{equation}\label{va-ep}
   \varepsilon= \left((m+E)^{2}\frac{v^{2}}{k^{2}_d}-(m^{2}-E^{2})\right).
\end{equation}
Using (\ref{o1}) and (\ref{o2}), we can construct the partner Hamiltonians $H_1$ and $H_2$
\begin{equation}\label{h1-h2}
    H_1=A^{-}_{0}A^{+}_{0}, ~~~~~H_2=A^{+}_{0}A^{-}_{0}
\end{equation}
or
\begin{eqnarray}
% \nonumber to remove numbering (before each equation)
  H_1 &=& -\frac{d^{2}}{dr^{2}}+\frac{k_{d}(k_{d}+1)}{r^{2}}-2(m+E)\frac{v}{r}+(m+E)^{2}\frac{v^{2}}{k^{2}_{d}}\label{h1} \\
  H_2 &=& -\frac{d^{2}}{dr^{2}}+\frac{k_{d}(k_{d}-1)}{r^{2}}-2(m+E)\frac{v}{r}+(m+E)^{2}\frac{v^{2}}{k^{2}_{d}}.\label{h2}
\end{eqnarray}
By comparing Eqs.~(\ref{h1}) and (\ref{h2})  with Eq.~(\ref{v12}),  it is clear that $a_1$ and $a_2$ are given by $a_1=k_d$ and $a_2=k_d+1$. By using (\ref{e}), (\ref{o1}) and (\ref{o2}), we find
\begin{equation}\label{en}
   E^{2}_{n}-m^{2}+\frac{v^{2}}{k^{2}_d}(E_{n}+m)^{2}= (m+E_{n})^{2} v^{2}\left(\frac{1}{k^{2}_d}-\frac{1}{(k_d+n)^{2}}\right),
\end{equation}
and the entire spectrum can be obtained as
\begin{equation}\label{en1}
    E^{(2)}_{n}=E_n = m\frac{1-\frac{v^{2}}{(k_d+n)^{2}}}{1+\frac{v^{2}}{(k_d+n)^{2}}}, ~~~n=0,1,2,...
\end{equation}
Eq.(\ref{E1}) shows that $A^{+}_{0} A^{-}_{0}$ and $A^{-}_{0}A^{+}_{0} $ have the same spectrum except when $A^{-}_{0}\tilde{\psi}_{2}=0$. Thus, the ground-state wave function becomes
\begin{equation}\label{gs}
    \tilde{\psi}^{(2)}_{0}= r^{k_d} e^{-(m+E)\frac{v}{k_d}r}
\end{equation}
and the ground- state energy is
\begin{equation}\label{grstate}
    E^{(2)}_{0}=m\frac{1-\frac{v^{2}}{k^{2}_{d}}}{1+\frac{v^{2}}{k^{2}_{d}}}.
\end{equation}
Thus, the first order differential equation (\ref{o1}) implies
\begin{equation}\label{psi10}
    \tilde{\psi}^{(1)}_{0}=r^{-k_d} e^{\frac{(m+E)v r}{k_d}}\int^{r} dz~ z^{2k_d}~ e^{-\frac{2(m+E)v z}{k_d}}.
\end{equation}
First excited state of $\tilde{\psi}_2$ is given by $\tilde{\psi}^{(2)}_1 \propto A^{+}_0 \tilde{\psi}^{(2)}_0$. The complete set of solutions can be obtained by using (\ref{suk}) and $\sqrt{m^{2}-E^{2}}=(m+E)v/(k_d+n)$ from the energy relation (\ref{en}):
\begin{equation}\label{wf1}
    \tilde{\psi}^{(2)}_{n}\propto r^{k_d}~ e^{-\frac{(m+E)v}{k_d}r}~\mathcal{L}^{2k_d-1}_{n}\left(2 \frac{(m+E)v}{k_d}r\right)
\end{equation}
where $\mathcal{L}^{b}_{n}(x)$ are Laguerre polynomials.

In order to select square integrable wave functions, we have to distinguish different cases. A solution of the differential equation $A^{-}_0\tilde{\psi}_{2}=0$ yields (\ref{gs}) with the ground state energy (\ref{grstate}). We see that $\tilde{\psi}^{(2)}_{0}$ satisfies the boundary conditions when $k_d > 0$, $v > 0$. However, if we now look at (\ref{psi10}), we see that the $n=0$ solution for  $\tilde{\psi}_{1}$ does not satisfy the boundary condition when $k_d > 0$, $v > 0$. Thus, we see that $E=E_{0}$ is not an eigenvalue of $H_1$ because $H_1$ has no corresponding $L^{2}$ solution. This means that when spin symmetry occurs, it is not possible for $n=0$ to construct a  normalisable spinor in terms of $\tilde{\psi}_{1}$ and $\tilde{\psi}_{2}$. On the other hand, in the Dirac equation for a central Coulomb field problem, without the restriction of spin symmetry, one can indeed find a singlet state when $n=0$, for example if $S = 0$~\cite{sukumar}.
%%%%%%%%%%%%%%%%%%%%%%%%%%%%%%%%%%%%%%%%%%%%%%%%%%%%%%
\subsection{Pseudo-spin-symmetric problems }
%%%%%%%%%%%%%%%%%%%%%%%%%%%%%%%%%%%%%%%%%%%%%%%%%%%%%%
In this case, the matrix equation becomes
\begin{eqnarray}\label{m2}
\left(
  \begin{array}{c}
    \psi'_1 \\    \psi'_2 \\  \end{array} \right)+\left(   \begin{array}{cc}
     \frac{k_d}{r} & 2V(r) \\      0 & -\frac{k_d}{r} \\    \end{array}
 \right)  \left(   \begin{array}{c}     \psi_1 \\     \psi_2 \\   \end{array} \right) =\left(
    \begin{array}{cc}      0 & m+E \\       m-E & 0 \\
    \end{array}  \right)  \left(    \begin{array}{c}       \psi_1 \\       \psi_2 \\    \end{array}  \right).
\end{eqnarray}
Again, the matrix which diagonalizes the matrix with terms $\frac{1}{r}$ in (\ref{m2}) is given by
\begin{eqnarray*}
D=\left(
  \begin{array}{cc}
    1 & \frac{v}{k_d} \\    0 & 1 \\  \end{array}\right).
\end{eqnarray*}
If $D^{-1}$ is applied to (\ref{m2}) from the left, as above, then we obtain
\begin{equation}\label{o3}
    A^{+}_{0} \tilde{\psi_{1}}=\left(m+E-\frac{v^{2}}{k^{2}_d}(m-E)\right)\tilde{\psi_{2}}
\end{equation}
and
\begin{equation}\label{o4}
    A^{-}_{0}\tilde{\psi_{2}}=(E-m)\tilde{\psi_{1}},
\end{equation}
where the operators can be defined as
\begin{equation}\label{op1}
    A^{\pm}_{0}=\pm\frac{d}{dr}+\frac{k_d}{r}-(E-m)\frac{v}{k_d}.
\end{equation}
For this case, the eigenvalue equations are
\begin{eqnarray}\label{E2}
% \nonumber to remove numbering (before each equation)
  A^{-}_{0} A^{+}_{0} \tilde{\psi_{1}} &=& \varepsilon \tilde{\psi_{1}}\\
  A^{+}_{0} A^{-}_{0}  \tilde{\psi_{2}} &=& \varepsilon \tilde{\psi_{2}}, \nonumber
\end{eqnarray}
where
\begin{equation}\label{ep}
    \varepsilon=\left(E^{2}-m^{2}+\frac{v^{2}}{k^{2}_d}(E-m)^{2}\right).
\end{equation}
Now the partner Hamiltonians become
\begin{eqnarray}
% \nonumber to remove numbering (before each equation)
  H_1 &=& -\frac{d^{2}}{dr^{2}}+\frac{k_d (k_d+1)}{r^{2}}-2(E-m)\frac{v}{r}+(E-m)^{2}\frac{v^{2}}{k^{2}_d} \\
  H_2 &=&  -\frac{d^{2}}{dr^{2}}+\frac{k_d (k_d-1)}{r^{2}}-2(E-m)\frac{v}{r}+(E-m)^{2}\frac{v^{2}}{k^{2}_d}.
\end{eqnarray}
The parameters are again $a_1=k_d$ and $a_2=k_d+1$. By using (\ref{ep}), we can obtain the spectrum for this case as
\begin{equation}
    E^{(2)}_{n}=-m\frac{1-\frac{v^{2}}{(k_d+n)^{2}}}{1+\frac{v^{2}}{(k_d+n)^{2}}}.
\end{equation}
From (\ref{o3}) and (\ref{o4}), we have $A^{-}_{0} \tilde{\psi}^{0}_{2}=0$, thus
\begin{equation}\label{p2}
    \tilde{\psi}^{(2)}_{0} = r^{k_d} e^{-(E-m)\frac{v}{k_d}r} , ~~~~~~~\tilde{\psi}^{(1)}_{0}=0
\end{equation}
can be obtained. If we look at $\tilde{\psi}^{(2)}_{0}$ in (\ref{p2}), it is clear that the positive values for $k_d$ and negative values for $v$ must be used in order to obtain normalisable solutions $ \tilde{\psi}^{(2)}_{0}$. For $n=0$, $v <0 $ we have normalisable solutions $\tilde{\psi}^{(1)}_{0}, ~~ \tilde{\psi}^{(2)}_{0}$. And the  solutions $\tilde{\psi}^{(2)}_{n}$ are given by
\begin{equation}\label{ent}
    \tilde{\psi}^{(2)}_{n} \propto r^{k_d}  e^{-\frac{(E-m)v}{k_d}r}~\mathcal{L}^{2k_d-1}_{n}\left(2 \frac{(E-m)v}{k_d}r\right).
\end{equation}
%%%%%%%%%%%%%%%%%%%%%%%%%%%%%%%%%%%%%%%%%%
\section{The Kratzer Potential}
%%%%%%%%%%%%%%%%%%%%%%%%%%%%%%%%%%%%%%%%%%
We shall use the Coulomb problem of the previous section as a guide in our analysis of the Kratzer case.
We write the Kratzer potential\cite{kratzer} in the form:
\begin{equation}\label{kr}
    V(r)=\frac{\lambda}{r^{2}}-\frac{v}{r}+c.
\end{equation}
%%%%%%%%%%%%%%%%%%%%%%%%%%%%%%%%%%%%%%%%%
\subsection{Spin symmetric case, $S=V$}
%%%%%%%%%%%%%%%%%%%%%%%%%%%%%%%%%%%%%%%%%
In the spin symmetric case, we can give the eigenvalue equation by using (\ref{eq3}) and (\ref{eq4}):
\begin{equation}\label{kr-e}
    \left(-\frac{d^{2}}{dr^{2}}+\frac{k_d(k_d+1)+2\lambda(E+m)}{r^{2}}-
    2(E+m)\frac{v}{r}+2(E+m)c\right)\psi_{1}=(E^{2}-m^{2})\psi_{1}.
\end{equation}
Now, we introduce the operators $A^{\pm}_{0}$ as
\begin{equation}\label{pm}
    A^{\pm}_{0}=\pm \frac{d}{dr}+\beta_1(r).
\end{equation}
Using the Coulomb case of section~3 as a guide, we adopt an ansatz for $\beta_1(r)$ of the form
\begin{equation}\label{beta}
    \beta_1(r)=\frac{s_{1}}{r}-\frac{\alpha_1}{s_{1}},
\end{equation}
where $s_1$ and $\alpha_1$ are the positive constants. Partner Hamiltonians are given by
\begin{equation}\label{Hh1}
    H_1=A^{-}_0 A^{+}_0=-\frac{d^{2}}{dr^{2}}+\frac{s_1(s_1+1)}{r^{2}}-\frac{2\alpha_1}{r}+\frac{\alpha_1^{2}}{s^{2}_1}
\end{equation}
and
\begin{equation}\label{Hh2}
    H_2=A^{+}_0 A^{-}_0=-\frac{d^{2}}{dr^{2}}+\frac{s_1(s_1-1)}{r^{2}}-\frac{2\alpha_1}{r}+\frac{\alpha_1^{2}}{s^{2}_1}.
\end{equation}
The relation between the potentials is given by
\begin{equation}\label{r}
    V_1(r,s_1;v)=V_2(r,s_1+1;v)+\frac{\alpha_1^{2}}{s^{2}_1}-\frac{\alpha_1^{2}}{(s_1+1)^{2}}.
\end{equation}
Here, we present the expressions for the first-order operators as
\begin{equation}\label{op12}
    A^{+}_{0} \psi_{1}=\mu \psi_{2}, ~~~~~~A^{-}_{0} \psi_{2}=\nu \psi_{1}.
\end{equation}
If we compare  (\ref{Hh1})-(\ref{r}) and  (\ref{kr-e}), we obtain $s_{1}$, $\alpha_1$, and the product $\mu \nu$ (to be factorised later):
\begin{equation}\label{cons}
    s_1=\frac{1}{2}+\sqrt{(k_d+\frac{1}{2})^{2}+2\lambda(E+m)}, ~~\alpha_1=v(E+m),
\end{equation}
and
\begin{equation}
\mu \nu=E^{2}-m^{2}-2(E+m)c+\frac{v^{2}(E+m)^{2}}{s^{2}_1}.
\end{equation}
Using (\ref{e}) and (\ref{r}), we have
\begin{equation}\label{muv}
    (\mu\nu)_n=v^{2}(E_{n}+m)^{2}\left(\frac{1}{s^{2}_1}-\frac{1}{(s_1+n)^{2}}\right).
\end{equation}
This leads to the following general energy relation, which is consistent with results obtained by different methods and reported in \cite{HY},
\begin{equation}\label{energy}
    (n+1/2+\sqrt{(k_d+1/2)^{2}+2\lambda(E_n+m)}) \sqrt{m^{2}-E^{2}_n+2c(E_n+m)}=v(E_n+m).
\end{equation}
Now we factorize the $\mu\nu$ expression in the form
\begin{eqnarray}
% \nonumber to remove numbering (before each equation)
  \mu &=& m+E \\
  \nu &=& E-m-2c+\frac{v^{2}}{s^{2}_1}(E+m).
\end{eqnarray}
The ground state wavefunction then reads
\begin{equation}\label{gs-kr}
    \psi^{(2)}_{0}=r^{s_1} e^{-\frac{v(m+E)r}{s_1}}
\end{equation}
and the ground state energy is
\begin{equation}\nonumber
    E_{0}=m\frac{1-\frac{v^{2}}{s^{2}_1}}{1+\frac{v^{2}}{s^{2}_1}}+\frac{2c}{1-\frac{v^{2}}{s^{2}_1}}.
\end{equation}
And the solutions can be obtained as
\begin{equation}\label{sol}
    \psi^{(2)}_n \propto r^{s_1}  e^{-\frac{(m+E)v}{s_1}r}~\mathcal{L}^{2s_1-1}_{n}\left(2 \frac{(m+E)v}{s_1}r\right).
\end{equation}
The boundary conditions imply $v>0$. Again, $\psi^{(1)}_{0}$ does not satisfy the boundary condition of square-integrability.
%%%%%%%%%%%%%%%%%%%%%%%%%%%%%%%%%%%%%%%%%%%%%%%%%%%
\subsection{Pseudo-spin symmetric case, $S=-V$}
%%%%%%%%%%%%%%%%%%%%%%%%%%%%%%%%%%%%%%%%%%%%%%%%%%%
For $S=-V$, we can give the eigenvalue equation by using (\ref{eq3}) and (\ref{eq4}):
\begin{equation}\label{kr-e1}
    \left(-\frac{d^{2}}{dr^{2}}+\frac{k_d(k_d-1)+2\lambda(E-m)}{r^{2}}-
    2(E-m)\frac{v}{r}+2(E-m)c\right)\psi_{2}=(E^{2}-m^{2})\psi_{2}.
\end{equation}
And we write the ansatz for $\beta_2(r)$ as;
\begin{equation}\label{beta}
    \beta_2(r)=\frac{s_{2}}{r}-\frac{\alpha_2}{s_{2}}.
\end{equation}
This leads to partner Hamiltonians $H_1$ and $H_2$ given by:
\begin{equation}\label{Hh1s}
    H_1=A^{-}_0 A^{+}_0=-\frac{d^{2}}{dr^{2}}+\frac{s_2(s_2+1)}{r^{2}}-\frac{2\alpha_2}{r}+\frac{\alpha_2^{2}}{s^{2}_2}
\end{equation}
\begin{equation}\label{Hh2p}
    H_2=A^{+}_0 A^{-}_0=-\frac{d^{2}}{dr^{2}}+\frac{s_2(s_2-1)}{r^{2}}-\frac{2\alpha_2}{r}+\frac{\alpha_2^{2}}{s^{2}_2}.
\end{equation}
We again compare (\ref{kr-e1}) and (\ref{Hh2p}) and obtain:
\begin{equation}\label{cons2}
    s_2=\frac{1}{2}+\sqrt{(k_d-1/2)^{2}+2\lambda(E-m)},~~\alpha_2=v(E-m).
\end{equation}
Following the same steps as in the spin symmetric case, the energy relation is found to be
\begin{equation}\label{enpss}
    v(E_n-m)=(n+1/2+\sqrt{(k_d-1/2)^{2}+2\lambda(E_n-m)})\sqrt{2c(E_n-m)+m^{2}-E_n^{2}}
\end{equation}
and the ground state wavefunction is
\begin{equation}\label{g-s-2}
    \psi^{(2)}_0=r^{s_2} e^{-(E-m)\frac{v}{s_2}r}
\end{equation}
and the same conclusion is valid for (\ref{g-s-2}) as given in the Coulomb potential case. Here, if $v<0$ we have normalisable solutions for the ground state with $\psi^{(1)}_{0}=0$ and $ \psi^{(2)}_0$ given in (\ref{g-s-2}). Thus, solutions are given by
\begin{equation}\label{sol2}
    \psi^{(2)}_n \propto r^{s_2} e^{-(E-m)\frac{v}{s_2}r} ~\mathcal{L}^{2s_2-1}_{n}\left(2 \frac{(E-m)v}{s_2}r\right).
\end{equation}
\section{Shift operators and the Kratzer potential}
 Introducing an operator $O_E$,
\begin{equation}\label{o12}
    O_E=-r^{2} \frac{d^{2}}{dr^{2}}+(2c(E+m)-E^{2}+m^{2})r^{2}-2 v (E+m)r
\end{equation}
we can express (\ref{kr-e}) in the form
\begin{equation}\label{o-e}
    O_E ~ \psi_{E k_d}=- (k_d(k_d+1)+2\lambda(E+m))\psi_{E k_d}.
\end{equation}
In order to factorize (\ref{o-e}), let us first use $a=\frac{1}{(E+m) v}$ and $2c-E+m=\frac{1}{an^{2}}$. Using these parameters, we have
\begin{equation}\label{n}
    n=v \left(\frac{E+m}{2c-E+m}\right)^{\frac{1}{2}}.
\end{equation}
We observe that, (\ref{n}) can be obtained from (\ref{grstate}). For the bound states for this potential, we introduce an operator $O_n$ \cite{lange}
\begin{equation}\label{opC}
    O_{n}=-r^{2}\frac{d^{2}}{dr^{2}}-\frac{2}{a}r+\frac{r^{2}}{a^{2}n^{2}},
\end{equation}
\begin{equation}\label{opb}
    O_n+R_{\pm}=Q^{\mp}_{n\pm1}Q^{\pm}_{n},
\end{equation}
where $R_{\pm}$ are constants. Our aim is to construct the $Q^{\pm}_n$ in order to factorize $O_n.$  We assume an ansatz for $Q^{\pm}_n$, namely
\begin{equation}\label{a1}
    Q^{\pm}_n=\pm n z \frac{d}{dz}-\frac{z}{a}+n, ~~~~z=\frac{r}{n}, ~~~~ [z,p_z]=i.
\end{equation}
If we insert (\ref{a1}) in (\ref{opb}), we obtain $R_{\pm}=n(n\pm1)$. However, by use of (\ref{a1}) $Q^{\mp}_{n\pm1}Q^{\pm}_{n}$  leads to
\begin{equation}\label{oo}
    O_n+ n(n\pm 1)=\left(\mp r \frac{r}{dr}-\frac{r}{an}+n\pm 1\right)\left(\pm r\frac{d}{dr}-\frac{r}{an}+n\right).
\end{equation}
This means that we have to find an operator $D^{\pm}_n$ such that \cite{lange}
\begin{equation}\label{dop}
    D^{\pm}_n r=r\frac{n}{n\pm1}D^{\pm}_n, ~~~~~ D^{\pm} p_r =\frac{n\pm1}{n} p_r D^{\pm}_n, ~~~n\neq 1
\end{equation}
and
\begin{equation}\label{prop}
   (D^{\pm}_n)^{\dag}D^{\pm}_n=1, ~~~~(D^{\pm}_n)^{\dag}=D^{\mp}_{n\pm1}.
\end{equation}
With the aid of $D^{\mp}_{n\pm1}D^{\pm}_n=1$, one obtains
\begin{equation}\label{aid}
    Q^{\mp}_{n\pm1} Q^{\pm}_n=O_n+n(n\pm1),~~~~ n\neq1.
\end{equation}
Here, by using $D^{\pm}_n$,  $Q^{\pm}_n$ can be written as
\begin{equation}\label{v}
    Q^{\pm}_n=D^{\pm}_n T^{\pm}_n, ~~~~~~~ T^{\pm}_n=\pm r\frac{d}{dr}-\frac{r}{an}+n.
\end{equation}
From  (\ref{aid}), we conclude that $D^{\pm}_n ~|n \ell >=\epsilon^{\pm} ~ |n\pm 1, \ell >$.
By using the normalization condition, we get
\begin{equation}\label{norm}
   | \epsilon^{\pm}|^{2}=< n\ell | (T^{\pm}_n)^{\dag}T^{\pm}_n |n\ell >.
\end{equation}
From (\ref{o12}), (\ref{o-e}) and \cite{lange}, we see that
\begin{equation}\label{t}
    (T^{\pm}_n)^{\dag}=T^{\mp}_n \mp 1,~ <r>=\frac{a}{2}(3n^{2}-\ell(\ell+1)),~< r p_r>=\frac{i}{2},
\end{equation}
and in turn we obtain
\begin{equation}\label{eps}
    \epsilon^{\pm}=\sqrt{\frac{(n\pm1)}{n}(n(n\pm1)-\ell(\ell+1))}.
\end{equation}
It is clear that the minimum value for $n$ is $n_{min}=\ell+1$ from (\ref{eps}) and $Q^{-}_1 |10>=0$. Thus, if we apply $Q^{+}_{\ell+1} Q^{+}_{\ell+2}...|\ell+1, \ell>$, we get
\begin{equation}\label{nodes}
    n=N+\ell+1, ~N=0,1,...
\end{equation}
If we use (\ref{nodes}), (\ref{n}) and $k_d(k_d+1)+2\lambda(E+m)=\ell(\ell+1)$, we can obtain the energy relation of the Kratzer potential (\ref{energy}) for the spin symmetric case. Similar steps can, of course, be used for the corresponding pseudo-spin-symmetric problem.
%%%%%%%%%%%%%%%%%%%%%%%%%%%
\section{Conclusion}
%%%%%%%%%%%%%%%%%%%%%%%%%%%
In this paper we have used shape-invariant techniques to effect a supersymmetric analysis of the Dirac equation with the Coulomb and Kratzer potentials, for both the spin-symmetric ($S=V$) and pseudo-spin-symmetric ($S=-V$) cases. It is shown that the state $\tilde{\psi}^{(1)}_0$ is  missing for the spin symmetric case;  this does not happen with pseudo-spin symmetry. Thus the spin-symmetric and pseudo-spin-symmetric cases are qualitatively different.  For bound states, with either of these potentials, we find that the Coulomb coupling $v$ must have the appropriate sign: for spin symmetry, $v >0$; for pseudo-spin symmetry, $v<0$. This is consistent with the results of Ref.~\cite{HY}. We have shown that the factorization of the Dirac equation for the Kratzer potential can be performed by following similar steps to those used for the Coulomb case.  To illustrate the significance of this important alternative algebraic approach, we have employed the raising and lowering shift operators to obtain the spectrum generated by the Kratzer potential for the spin-symmetric case.

%%%%%%%%%%%%%%%%%%%%%%%%%%%%%%%
\section*{Acknowledgements}
%%%%%%%%%%%%%%%%%%%%%%%%%%%%%%%

One of us (RLH) gratefully acknowledges partial financial support
of this research under Grant No.\ GP3438 from the Natural Sciences
and Engineering Research Council of Canada; and one of us (\"OY) would like
to thank the Department of Mathematics and Statistics of Concordia University
 for its warm hospitality.

\medskip

%%%%%%%%%%%%%%%%%%%%%%%%%%%%%%

%%%%%%%%%%%%%%%%%%%%%%%%%%%%%%

\end{document}